\documentclass[12pt,a4paper,twoside]{article}
\usepackage{amssymb}
\usepackage{pdfsync}
\usepackage{times}
\usepackage{url}
\usepackage{imnvol}
\usepackage{tikz}

\usepackage[body={14.5cm,20.4cm}]{geometry}
\usepackage{amssymb}
\usepackage{hyperref}
\usepackage{enumerate}
\usepackage{enumitem}
\definecolor{gray50}{gray}{0.50}
\definecolor{gray75}{gray}{0.75}
\definecolor{gray85}{gray}{0.85}
\definecolor{gray90}{gray}{0.90}
\newcommand*{\mailto}[1]{\href{mailto:#1}{\nolinkurl{#1}}}

\newcommand{\doi}[1]{\href{http://dx.doi.org/#1}{doi: #1}}

\newtheorem{example}{Example}

\newcommand{\hH}{{\mathcal H}}

\newcommand{\R}{{\mathbb R}}
\newcommand{\N}{{\mathbb N}}
\newcommand{\Z}{{\mathbb Z}}
\newcommand{\C}{{\mathbb C}}

\newcommand{\E}{\mathrm{e}}
\newcommand{\I}{\mathrm{i}}
\newcommand{\sgn}{\mathrm{sgn}}

\newcommand{\im}{\mathrm{Im}}

\newcommand{\x}{\mathrm{x}}

\newcommand{\loc}{\mathrm{loc}}

\newcommand{\oo}{o}

\newcommand{\cS}{\mathcal{S}}

\newcommand{\dip}{\upsilon}

\newcommand{\ledot}{\,\cdot\,}
\newcommand{\redot}{\cdot\,}

\newcommand{\be}{\begin{equation}}
\newcommand{\ee}{\end{equation}}

\newcommand{\wt}{\widetilde}

\begin{document}
\def\Volume{233}
\def\Jahrgang{70}
\month 12
\year 2016

\title{The Camassa--Holm Equation and The String Density Problem}
\author{Jonathan Eckhardt}
\author{Aleksey Kostenko}
\author{Gerald Teschl}

\Artikel
	The Camassa--Holm Equation \neuezeile
	and The String Density Problem \neuezeile
	~;
	Jonathan Eckhardt, Aleksey Kostenko and Gerald Teschl;
	Universit\"at Wien;

The Camassa--Holm (CH) equation 
\begin{equation}\label{eqnCH}
   u_{t} -u_{xxt} +2\kappa u_x = 2u_x u_{xx} - 3u u_x + u u_{xxx},
\end{equation} 
is an extensively studied nonlinear equation. It first appeared as an abstract bi-Hamiltonian partial differential equation in an article of Fuchssteiner and Fokas \cite{fofu81} but did not receive much attention until Camassa and Holm \cite{caho93} (see also \cite{chh94}) derived it as a nonlinear wave equation that models unidirectional wave propagation on shallow water and discovered its rich mathematical structure. In this context $u(x,t)$ represents the fluid velocity in the $x$ direction at time $t$ and the real constant $\kappa$ is related to the critical shallow water wave speed. Regarding the hydrodynamical relevance, we refer to the more recent articles \cite{cola09, io07, jo02}. Apart from this, the CH equation was also found in \cite{dai} as a model for nonlinear waves in cylindrical hyperelastic rods.

Since its discovery, the literature on the CH equation has been growing exponentially (at the moment, the paper by Camassa and Holm \cite{caho93} has more than 2400 citations in Google Scholar, 1698 in Scopus and 1016 in MathSciNet) and it is impossible to give a comprehensive overview here.  In fact, our main focus in this review paper lies on understanding the CH equation and its so-called conservative solutions via the inverse scattering transform (IST) approach. It was noticed by Camassa and Holm that the CH equation is completely integrable in the sense that it enjoys a Lax pair structure and hence may be treated with the help of the IST approach in principal. The corresponding isospectral problem is a Sturm--Liouville problem of the form 
  \be\label{eq:Spec}
  -y'' + \frac{1}{4}y=z\, y\, \omega(\ledot,t),
  \ee
 where $\omega = u - u_{xx} +\kappa$ is known as the {\em momentum}. Using a simple change of variables (and dropping the time dependence for a moment), this spectral problem can be transformed into the spectral problem for an inhomogeneous string 
  \be\label{eq:spstring}
  -f'' =z\, f\, \wt{\omega}.
  \ee
 In the 1950's,  M.\ G.\ Krein developed direct and inverse spectral theory for such strings, assuming that $\wt{\omega}$ is a nonnegative, locally finite measure on $[0,L)$ for some $L\in (0,\infty]$. Of course, the original motivation of Krein was far from applications to nonlinear equations (see, for instance, \cite[Appendix 3]{gg}). Unlike in the case for strings, applications to the CH equation make it necessary to deal not only with nonnegative but also with real-valued (signed) Borel measures, that is, with {\em indefinite strings}. 
 
 The structure of the present review is as follows: In the next section we give a brief historical account on the CH equation. Section \ref{sec:MP} then discusses multi-soliton solutions for (\ref{eqnCH}) in the case when $\kappa=0$; the so-called multi-peakons. In the next two sections we touch upon the concepts of global conservative solutions and generalized indefinite strings. In Section~\ref{sec:integrability}, we overview recent progress in the understanding of the conservative CH flow as a completely integrable nonlinear flow. In the final section we provide an account on the long-time behavior of solutions to the CH equation.

\section{The Camassa--Holm equation}\label{sec:hist}
     
\subsubsection*{Bi-Hamiltonian structure} In the rest of our paper, we will be mostly concentrated on the Cauchy problem for the CH equation on the line, that is, $(x,t)\in \R\times\R_+$ (note that
negative times are covered by the transformation $(x,t)\to (-x,-t)$, which leaves (\ref{eqnCH}) invariant). Then the CH equation can be written in the equivalent form
\begin{eqnarray}\label{eqnCHweak}
u_t + uu_x + p_x = 0,\qquad
p(x,t)=\frac{1}{2} \int_{\R} \E^{-|x-y|}\left( 2\kappa u + u^2 + \frac{1}{2} u_x^2\right)dy,
\end{eqnarray}
which is reminiscent of the 3-D incompressible Euler equations. The original derivation of (\ref{eqnCH}) was obtained by approximating directly in the Hamiltonian for Euler's equations in the shallow water regime (see \cite{caho93, chh94}) and hence the CH equation inherits the Hamiltonian structure
\be\label{eqnHam1}
m_t = -\Big(2\kappa \partial +m\partial + \partial m\Big)\frac{\delta \hH_1[m]}{\partial m}, \qquad m=u-u_{xx},
\ee
where the Hamiltonian is given by 
\be\label{eqnH1}
\hH_1[m] := \frac{1}{2}\int_{\R} u^2 + u_x^2\, dx.
\ee
This, in particular, implies that the $H^1$ Sobolev  norm of $u$ is a conserved quantity. In fact, the CH equation is bi-Hamiltonian \cite{fofu81, caho93}. Namely, it can also be written in the following alternative form:
\be\label{eqnHam2}
m_t = - (\partial -\partial^3)\frac{\delta \hH_2[m]}{\partial m}, \qquad \hH_2[m] := \frac{1}{2}\int_{\R} u^3 + uu_x^2 + 2\kappa u^2\, dx.
\ee
The latter leads to an infinite number of conserved quantities $\hH_n[m]$, $n\in\Z$, which are defined recursively by 
\[
(\partial -\partial^3)\frac{\delta \hH_{n+1}[m]}{\partial m} = \Big(2\kappa \partial +m\partial + \partial m\Big)\frac{\delta \hH_n[m]}{\partial m},\quad n\in\Z.
\]
Schemes for the computation of $\hH_n$ can be found in \cite{fish99, iva06, len05, rey02}.

\subsubsection*{Geometric formulation} Equation (\ref{eqnCH}) with $\kappa=0$ can be interpreted as the geodesic flow on the group of diffeomorphisms of the line with the Riemannian structure induced by the $H^1$ right-invariant metric. This resembles the fact that the Euler equation is an expression of the geodesic flow in the group of incompressible diffeomorphisms (see \cite{arn66, ebma70}).  
Equation (\ref{eqnCH}) represents the equations of motion in Eulerian coordinates, while the geometric interpretation corresponds to rewriting (\ref{eqnCH}) in Lagrangian coordinates. This connection turned out to be very useful in the qualitative analysis of solutions of the CH equation. 

For $\kappa\neq 0$, the CH equation represents the equation for geodesics on the Bott--Virasoro group \cite{mis98}. Let us also mention that the analogous correspondence for the Korteweg--de Vries (KdV) equation 
\be\label{eqnKdV}
u_t + u_{xxx} + 6uu_x = 0
\ee
was established in \cite{ovkh87} (see also \cite{khmi03}). In fact, there is a very close connection between the KdV and the CH equations. First of all, the Virasoro group (a one-dimensional extension of the group of smooth transformations of the circle) serves as  the symmetry group for these equations \cite{khmi03}. On the other hand, there is a Liouville correspondence between the CH and the KdV hierarchies \cite{mck03}, \cite{len04b}.

\subsubsection*{Dynamics of solutions} 
In this subsection we assume for simplicity that $\kappa=0$ (note that the transformation $u(x,t)\to u(x-\kappa t,t)+\kappa$
reduces (\ref{eqnCH}) to this case, but it does not preserve spatial asymptotics). 

The Sobolev spaces are the natural phase spaces for the CH equation since the Hamiltonian $\hH_1$ given by (\ref{eqnH1}) is exactly the $H^1$ norm of the solution at time $t$. One of the crucial differences between the CH and the KdV equations is the fact that the CH equation possesses both global solutions as well as solutions developing singularities in finite time. Moreover, the blow-up happens in a way which resembles wave breaking to some extent.

First of all, let us mention that the CH equation (\ref{eqnCH}) is locally well-posed in $H^s$ for any $s>3/2$ (the first result was obtained by Escher and Constantin \cite{coes98} and for further improvements see \cite{liol00}, \cite{rod01}). The problem with global well-posedness stems from the fact that the Sobolev norms $H^s$ are not controlled by the conservation laws if $s>1$ and hence one cannot extend local solutions automatically to the whole line and in fact, the blow-up can occur in finite time.  The singularity formation was first noticed by Camassa and Holm \cite{caho93}. Moreover, it was shown in \cite{coes98} that for any even initial data $u_0\in H^3(\R)$ with $u_0'(0)<0$ the corresponding solution does not exist globally. In particular, this result shows that initial data with arbitrary small $H^s$ norm may blow up in finite time. On the other hand, it was shown in \cite{coes98b} that the encountered blow-up looks as follows: the solution remains bounded but its slope becomes vertical in finite time, which resembles a breaking wave. Let us also mention that in certain situations it is possible to prove global existence. Namely, it was noticed in \cite{coes98} that solutions are global for $u_0\in H^s$ with $s>3/2$ if the corresponding momentum $\omega_0 = u_0 - u_0''$ is a positive finite measure. 

In contrast to classical solutions, weak solutions to the CH equation (\ref{eqnCHweak}) are global although they are not necessarily unique anymore. In \cite{xizh00}, Xin and Zhang proved the existence of global weak solutions for any  $u_0\in H^1(\R)$. It turns out that the positivity of the corresponding momentum plays a crucial role for the uniqueness of weak solutions. Namely, it was proved by Constantin and Molinet \cite{como00} that for $u_0 \in H^1(\R)$ such that the corresponding momentum $\omega_0$ is a positive finite measure on $\R$, a weak solution to (\ref{eqnCHweak}) exists and is unique for all times. Moreover, in this case $u$ is continuous with values in $H^1(\R)$ and the quantities $\hH_0$, $\hH_1$ and $\hH_2$ are conserved along the trajectories. In fact, the positivity of $\omega_0$ provides a criterion for the uniqueness of weak solutions: McKean \cite{mck98}, \cite{mck04} proved that a weak solution exists and is unique if $u_0 \in C^\infty(\R)\cap H^1(\R)$ is such that the set $S_- : = \{x\in\R:\ \omega_0(x)<0\}$ lies wholly to the right of the set $S_+ : = \{x\in\R:\ \omega_0(x)>0\}$. Hence, either the forward or backward CH flow blows up in finite time if both sets $S_+$ and $S_-$ have a nonzero Lebesgue measure. For further details and references we refer the reader to a survey by Molinet \cite{mol04}.

\subsubsection*{The Lax pair} The presence of infinitely many integrals of motion established in \cite{fofu81} indicates that the CH equation might be completely integrable. The latter was confirmed by Camassa and Holm \cite{caho93} by finding the corresponding Lax pair. Indeed, the CH equation can be formulated as the compatibility condition between 
\be\label{eq:specCH}
-y_{xx} +\frac{1}{4}y = z\, y\, \omega,\qquad \omega = u - u_{xx} + \kappa
 \ee 
and 
\be
y_t = \frac{1}{2} u_x y - \left(\frac{1}{2z} + u \right) y_x,
\ee
that is,
  $y_{xxt} = y_{txx}$
holds if and only if $u$ satisfies (\ref{eqnCH}). Let us also mention that the CH equation gives a counterexample to the Painlev\'e integrability test \cite{gipi95}. 

The spectral problem (\ref{eq:specCH}) is a  Sturm--Liouville problem. It very much resembles the 1-D Schr\"odinger spectral problem, $- y'' + q y = z\, y$,
which serves as the isospectral problem for the KdV equation (\ref{eqnKdV}). However, the spectral parameter $z$ is in the "wrong" place. Of course, under additional smoothness and positivity assumptions (for example $\kappa>0$, $\omega \in C^2(\R)$ and $\omega>0$ on $\R$), the Liouville transformation 
\be\label{eq:Liu1}
f(\x) = \omega(x)^{1/4}\, y(x),\quad \x(x) = x-\int^{\infty}_x \sqrt{\frac{\omega(s)}{\kappa}}-1\, ds,
\ee
converts (\ref{eq:specCH}) into the 1-D Schr\"odinger form
\be\label{eq:Liu2}
- f'' + Q(\ledot,t)f = z\, f,\quad Q(\x,t) = \frac{\kappa}{4}\frac{\omega_{xx}(x,t)}{\omega(x,t)^2} -\frac{\omega(x,t)-\kappa}{4\omega(x,t)}  - \frac{5\kappa}{16}\frac{\omega_{x}(x,t)^2}{\omega(x,t)^3}.
\ee
Hence, as for the KdV flow, in this case one can apply the well-developed inverse scattering theory for 1-D Schr\"odinger equations in order to integrate the CH flow using the IST approach (see, e.g., \cite{besasz98}, \cite{co01}, \cite{coiv06}, \cite{cogeiv06}). The same trick can be used to investigate the CH equation on the circle \cite{comc99}. However, the direct and inverse spectral theory for~(\ref{eq:specCH}) without these additional (positivity and smoothness) assumptions has not being developed and we postpone its further discussion to Section \ref{sec:strings}.

\subsubsection*{Solitons} One of the most interesting features of the CH equation is the presence of solitons and the simplicity of their interaction when $\kappa=0$ (
see Section \ref{sec:MP}). For $\kappa>0$, solitons are smooth but there is no closed form even for a one-soliton solution of the CH equation. However, the Liouville correspondence (\ref{eq:Liu1})--(\ref{eq:Liu2}) allows to obtain a detailed description of multi-soliton solutions. Namely, as for the KdV equation, multi-soliton profiles are reflectionless potentials for (\ref{eq:Liu2}) and one can employ this fact and the Liouville transform in order to get various representations of multi-soliton solutions (see, e.g., \cite{jo03}, \cite{lizh04}, \cite{mat05}, \cite{sch98}). Unfortunately, it is a difficult task to invert (\ref{eq:Liu1})--(\ref{eq:Liu2}) and this fact (at least partially) explains the absence of a closed form for multi-soliton solutions.


Finally, notice that the one-soliton solution is a traveling wave and a complete description of all weak traveling wave solutions to the CH equation (peakons, cuspons, stumpons, etc.) is given in \cite{len05b} (see also \cite{len04}).
 
\section{Multi-peakons and the moment problem}\label{sec:MP}

\subsubsection*{Soliton dynamics} In the dispersionless case $\kappa=0$, the traveling wave solution called {\em peakon} is given by 
\[ 
u(x,t) = p\, \E^{-|x-pt +c|}, 
\]
where $p$ and $c$ are real parameters.
It has a peak at $x_0(t) = pt-c$ and its height is equal to its speed (a positive peak travels to the right and a negative peak travels to the left). Since it obviously has a discontinuous first derivative at $x_0$, it has to be interpreted as a suitable weak solution of (\ref{eqnCHweak})  (see \cite{besasz00, coes98, como00, hora06}). 
It was noticed by Camassa and Holm \cite{caho93} that the multi-soliton solution to the CH equation with $\kappa=0$ is simply a linear combination of peakons
 \be\label{eqnMP}
  u(x,t) = \sum_{n=1}^N p_n(t)\, \E^{-|x-q_n(t)|}, 
 \ee
 where the coefficients $p_n$ and $q_n$  satisfy the  system of ordinary differential equations 
 \be\label{eqnMPsys}
  q_n'  = \sum_{k=1}^N p_k\, \E^{-|q_n-q_k|}, \qquad
  p_n'  = \sum_{k=1}^N p_n p_k\, \sgn(q_n - q_k)\, \E^{-|q_n-q_k|}.
 \ee
This system is Hamiltonian, that is,  
 \be\label{eqnMPsysH}
  \frac{d q_n}{dt} = \frac{\partial H(p,q)}{\partial p_n}, \qquad 
  \frac{d p_n}{dt} = - \frac{\partial H(p,q)}{\partial q_n}, 
 \ee
 with the Hamiltonian given by 
 \be\label{eq:H}
  H(p,q)= \frac{1}{2} \sum_{n,k=1}^N p_n p_k \, \E^{-|q_n-q_k|}=\frac{1}{4} \|u\|^2_{H^1(\R)}.
 \ee
 
 Before we proceed further, let us mention that the Hamiltonian system  (\ref{eqnMPsysH})--(\ref{eq:H}) is a special case of the Calogero--Fran\c{c}oise systems introduced in \cite{ca95,cafr96} 
 \be\label{eqnCF}
 H(p,q) = \frac{1}{2}\sum_{n,k=1}^N p_np_k G(q_n-q_k),\quad G(x) = a+b_+ \cos(\nu x) + b_-\sin(\nu|x|),
 \ee
 where $a$, $b_+$, $b_-$ and $\nu$ are arbitrary constants. Clearly, $a=0$, $b_+=1$ and $b_-=\nu=\I$ gives (\ref{eq:H}). Let us also mention that $a=0$, $b_+=\coth(1/2)$, $b_-=1$ and $\nu=\I$ gives rise to periodic multi-peakons (see \cite{bss05, comc99}); the limiting case $G(x) = a+b|x|+cx^2$ is associated to the Hunter--Saxton equation (see \cite{husa91, huzh94}).  
 
 The right-hand side in~(\ref{eqnMPsys}) is not Lipschitz if $q_n-q_k$ is close to zero and hence in this case, one cannot get existence and uniqueness of solutions of (\ref{eqnMPsys}) by using the standard arguments. 
 However, if we know in advance that all the positions stay distinct, then the right-hand side in (\ref{eqnMPsys}) becomes Lipschitz and thus the Picard theorem applies. Let us also mention that the Calogero--Fran\c{c}oise flows are completely integrable in the Liouville sense, that is, there exist $N$ integrals of motion in involution \cite{cafr96}. However,  the classical Arnold--Liouville theorem is not applicable since the Hamiltonians (\ref{eqnCF}) are not continuously differentiable whenever $b_-\neq 0$.
 
One of the most prominent features of multi-peakons is the fact that almost all qualitative properties of solutions to the CH equation can be seen just by considering multi-peakons, i.e., finite dimensional reductions (\ref{eqnMPsysH})--(\ref{eq:H}) of the infinite dimensional Hamiltonian system (\ref{eqnHam1})--(\ref{eqnH1}). For example, the behavior of multi-peakon solutions crucially depends on whether all the heights $p_n$ of the single peaks are of the same sign or not. Notice that the corresponding momentum is simply given by 
\be\label{eq:omegaMP}
\omega(\ledot,t) = 2 \sum_{n=1}^N p_n(t) \delta_{q_n(t)}. 
\ee    
So, $\omega$ is a positive measure precisely when all the heights are positive and in this case, all the positions $q_n$ of the peaks stay distinct, move to the right and the system~(\ref{eqnMPsys}) allows a unique global solution \cite{coes98, como00, hora06}. Otherwise, some of the positions $q_n$ of the peaks will collide eventually, which causes the corresponding heights $p_n$ to blow up in finite time \cite{caho93}. All this happens in such a way that the solution $u$ in~(\ref{eqnMP}) stays uniformly bounded but its derivative develops a singularity at the points where two peaks collide. Let us demonstrate this by considering the interaction of two peakons.
 
 \begin{example}[Two peakons]\label{ex:3.1} 
Consider the case $N=2$ 
and assume that $q_1(0) < q_2(0)$. 
Introducing the new variables $Q=q_2 - q_1$ as well as $P=p_2 - p_1$ and noting that $Q> 0$ in a vicinity of zero, we can rewrite the system as
\[
Q' = P (1-\E^{-Q}),\qquad P' = \frac{P_0^2-P^2}{2}\E^{-Q},
\] 
where $P_0 \equiv p_1(t)+p_2(t)$ is a constant of motion. 
Notice also that 
\be\label{eq:energy_2peaks}
2H(p,q) = \frac{P_0^2 + P^2}{2} +\frac{P_0^2 - P^2}{2}\E^{-Q} = P_0^2 - \frac{P_0^2 - P^2}{2}(1-\E^{-Q})\equiv 2H_0^2.
\ee
Therefore, we get
\be\label{eq:system_2peaks}
 P' = 2H_0^2-\frac{P_0^2+P^2}{2},\qquad Q' = 2P\frac{P_0^2 - 2H_0^2}{P_0^2 - P^2}.
\ee
One then easily obtains 
\be\label{eq:P_2peaks}
P(t) = h_0\frac{(P(0) + h_0)\E^{h_0t} + (P(0) - h_0)}{(P(0) + h_0)\E^{h_0t} - (P(0) - h_0)},\quad h_0 = \sqrt{4H_0^2 - P_0^2},
\ee
and 
\be\label{eq:Q_2peaks}
Q(t) =  Q(0) + \log\left| \frac{\left(\E^{h_0t} - \frac{P_0 + h_0}{P_0 - h_0}\frac{P(0) - h_0}{P(0) + h_0}\right)\left(\E^{h_0t} - \frac{P_0 - h_0}{P_0 + h_0}\frac{P(0) - h_0}{P(0) + h_0}\right)}{\left(1 - \frac{P_0 + h_0}{P_0 - h_0}\frac{P(0) - h_0}{P(0) + h_0}\right)\left(1 - \frac{P_0 - h_0}{P_0 + h_0}\frac{P(0) - h_0}{P(0) + h_0}\right)\E^{h_0t}}\right|.
\ee
Clearly, $P$ is discontinuous only if $(P(0) - h_0)(P(0) + h_0)>0$ and in this case  
\be\label{eq:tcrit}
P(t) \to -\infty \quad {\rm as}\quad t\to t^\times :=\frac{1}{h_0}\log\left(\frac{P(0) - h_0}{P(0) + h_0}\right). 
\ee
Notice that 
\[
(P(0) - h_0)(P(0) + h_0) = P(0)^2 +P_0^2 - 4H_0^2 = (P(0)^2 - P_0^2)\E^{-Q} 
\]
is positive only if $p_1(0)p_2(0)<0$. 
Hence there are two distinct cases. If $p_1(0)p_2(0)>0$, i.e., both peaks are of the same sign, then the solution is global and peaks never collide. If we have a peakon-antipeakon interaction, then the blowup happens at $t=t^\times$ given by~(\ref{eq:tcrit}) and in this case $P(t) = p_2(t) - p_1(t)$ tends to $-\infty$ and $Q(t)=q_2(t)-q_1(t)$ tends to zero as $t$ approaches $t^\times$.
 \end{example}
 
\subsubsection*{Complete integrability and the Stieltjes moment problem} It was noticed by Beals, Sattinger and Szmigielski \cite{besasz00} that similar to the finite Toda lattice on the line \cite{mos75}, the multi-peakon flow can be solved by using the solution of the Stieltjes moment problem \cite{sti}. More precisely, consider the corresponding spectral problem (\ref{eq:specCH}) 
 with the moment $\omega$ given by (\ref{eq:omegaMP}) (we omit the time dependence):
 \be\label{eq:specstr}
 -y'' + \frac{1}{4}y = z\, y\,\omega,\qquad \omega = 2\sum_{n=1}^N p_n\delta_{q_n}.
 \ee
Without loss of generality, we can assume that $p_n\neq 0$ for all $n\in \{1,2,...,N\}$ and 
\[ 
-\infty < q_1<q_2<...<q_N<\infty.
\]
 Some function $y$ is a solution of the differential equation~(\ref{eq:specstr}) if it satisfies 
 \be\label{eqnDEdiscr01}
   -y'' + \frac{1}{4} y = 0 
 \ee
 away from the points $\lbrace q_1,\ldots,q_N\rbrace$, together with the interface conditions  
 \be
  \left(\begin{array}{c} y(q_n+) \\ y'(q_n+) \end{array}\right)  =
\left(\begin{array}{cc} 1 & 0  \\  -2z p_n  & 1 \end{array}\right)
\left(\begin{array}{c} y(q_n-) \\ y'(q_n-) \end{array}\right),\quad n\in\{1,\dots, N\}.
\label{eqnDEdiscr02}
 \ee

The set of all values $z\in\C$ for which there is a nontrivial solution of the differential equation~(\ref{eq:specstr}) that lies in $H^1(\R)$ is referred to as {\em the spectrum} $\sigma$ of the spectral problem~(\ref{eq:specstr}). 
 Note that in this case, the solution in $H^1(\R)$ of this differential equation is unique up to scalar multiples. 
Since the measure $\omega$ has a compact support, for every $z\in\C$ one has spatially decaying solutions $\phi_\pm(z,\cdot\,)$ of~(\ref{eq:specstr}) with 
 $\phi_\pm(z,x) = \E^{\mp\frac{x}{2}} $
 for all $x$ near $\pm\infty$. 
 In particular, note that $\phi_\pm(\,\cdot\,,x)$ and $\phi_\pm'(\,\cdot\,,x)$ are real polynomials for each fixed $x\in\R$. 
 The Wronski determinant of these solutions  
\be\label{eqnWronski}
 W(z) = \phi_+(z,x) \phi_-'(z,x) - \phi_+'(z,x) \phi_-(z,x), \quad z\in\C
\ee
 is independent of $x\in\R$ and vanishes at some point $\lambda\in\C$ if and only if the solutions $\phi_-(\lambda,\cdot\,)$ and $\phi_+(\lambda,\cdot\,)$ are linearly dependent. As a consequence, one sees that the spectrum $\sigma$ is precisely the set of zeros of the polynomial $W$ and, moreover, the spectrum $\sigma$ of (\ref{eq:specstr}) consists of exactly $N$  real and simple eigenvalues (see, e.g., \cite{besasz00}).  Associated with each eigenvalue $\lambda\in\sigma$ is the quantity
  \be\label{eq:NC}
  \frac{1}{\gamma_{\lambda}}  := 
  \int_\R |\phi_-'(\lambda,x)|^2 dx+\frac{1}{4}\int_\R |\phi_-(\lambda,x)|^2 dx>0,  
  \ee
 which is referred to as the (modified) {\em norming constant} (associated with $\lambda$). 
 
 The central role in the inverse spectral theory for (\ref{eq:specstr}) is played by {\em the  Weyl--Titchmarsh $m$-function}, which is defined by
 \be\label{eq:Mmp}
M(z) = \lim_{x\to -\infty} -\frac{1}{z}\frac{W(\phi_+(z,x),\E^{-x/2})}{W(\phi_+(z,x),\E^{x/2})} =  \lim_{x\to -\infty} -\frac{W(\phi_+(z,x),\E^{-x/2})}{zW(z)} ,\quad z\in\C\setminus\R. 
 \ee
 $M$ is a Herglotz--Nevanlinna function and it admits 
 the partial fraction expansion:  
 \be\label{eqn:3.2}
 M(z) =  \sum_{\lambda\in\sigma} \frac{\gamma_\lambda}{\lambda-z}, \quad z\in\C\backslash\R.
 \ee

On the other hand, taking into account the fact that $\phi_+(z,\cdot)$ solves the difference equation (\ref{eq:specstr}), it is also possible to write down a finite continued fraction expansion for $M$ in terms of $\omega$: 
\be\label{eqnContFrac}
   zM(z) - 1 = \frac{1}{\displaystyle -l_{0} + \frac{1}{\displaystyle m_1\,z + \frac{1}{\displaystyle \;\ddots\; + \frac{1}{\displaystyle -l_{N-1} + \frac{1}{\displaystyle m_N\, z - \frac{1}{l_N}}}}}}\ , \qquad z\in\C\backslash\R,
  \ee
where 
 \be\label{eqnan}
m_n  = 8\, p_n   \cosh^2\left(\frac{q_n}{2}\right),\quad   l_n  =  \frac{1}{2} \left( \tanh\left(\frac{q_{n+1}}{2}\right) - \tanh\left(\frac{q_n}{2}\right)\right). 
  \ee
 Hereby, we set $q_0=-\infty$ and $q_{N+1}=\infty$ for simplicity of notation.
    
In the case when all $p_n$ are positive, the classical result of Stieltjes \cite{sti} (see also \cite[\S 13]{kakr74}) recovers the coefficients (\ref{eqnan}) in terms of the spectrum $\sigma$ and the corresponding norming constants $\{\gamma_\lambda\}_{\lambda\in\sigma}$. For that purpose, one needs to consider the Laurent expansion of $M(z)$ at infinity (taking into account the partial fraction expansion (\ref{eqn:3.2})):
\be\label{eqnMHN}
  M(z) - \frac{1}{z} = -\sum_{k=0}^{\infty} \frac{s_{k}}{z^{k+1}}, \quad |z|\rightarrow \infty;\qquad 
  s_k=  \left\{\begin{array}{cc} 
  1+ \sum_{\lambda\in\sigma}{\gamma_\lambda}, & k=0,\\[2mm]
  \sum_{\lambda\in\sigma}{\lambda^{k}}{\gamma_\lambda}, & k\in\N. \end{array}\right. 
  \ee
Introducing the Hankel determinants 
 \be\label{eqnDel01}
  \Delta_{0,k}  = \big| s_{i+j}\big|_{i,j=0}^k,
  \quad  
  \Delta_{1,k}  =  \big| s_{i+j+1}\big|_{i,j=0}^k,
  \ee
the formulas of Stieltjes read as follows
 \be\label{eqnCoanInv}
m_n =  \frac{\Delta_{0,n}^2}{\Delta_{1,n-1} \Delta_{1,n}}, \quad  l_{n-1} = \frac{\Delta_{1,n-1}^2}{\Delta_{0,n-1} \Delta_{0,n}},\quad n\in\{1,\dots,N\}. 
 \ee
Since $s_k$, $k\in\N$ are the moments of a nonnegative measure $\rho = \delta_0+\sum_{\lambda \in\sigma} \gamma_{\lambda}\delta_\lambda$,
the Hankel determinants $\Delta_{0,n}$ are positive for all $n\in\{0,\dots,N\}$ (see \cite{ga}, \cite{sti}). On the other hand, the Hankel determinants $\Delta_{1,n}$ are positive for all $n\in\{0,\dots,N\}$ if and only if the support of $\rho$ is contained in $[0,\infty)$, i.e., the spectrum $\sigma$ consists only of positive eigenvalues, which is further equivalent to the positivity of the measure $\omega$. Introducing the time dependence
\[
\dot{\gamma}_\lambda = \frac{1}{2\lambda} \gamma_\lambda, \quad \lambda\in\sigma,
\]
one then can integrate the multi-peakon flow by using the Stieltjes solution of the moment problem (\ref{eqnan})--(\ref{eqnCoanInv}) if all $p_n$ are positive.  
Moreover, the formulas (\ref{eqnContFrac})--(\ref{eqnCoanInv}) remain true until the denominators in (\ref{eqnCoanInv}) become zero, that is, one can exploit Stieltjes solution of the moment to integrate the multi-peakon flow in the general case (see \cite{besasz00} for further details). One of the important observations made in \cite{besasz00} is that one of the Hankel determinants $\Delta_{1,n}$ vanishes exactly when two adjacent peakons collide. 

\section{Generalized indefinite strings}\label{sec:strings}
 
The direct and inverse spectral theory for  the spectral problem (\ref{eq:specCH}) is of vital importance for investigating the CH equation by using the IST approach. Assume for a moment that $\omega$ is a locally finite measure on $\R$. A simple change of variables
\be\label{eq:LTstring}
x\mapsto \x = \x(x) = \frac{1}{2}\tanh(x/2), \quad x\in \R;\quad f(\x):=\frac{y(x)}{2\cosh(x/2)},
\ee
transforms the spectral problem (\ref{eq:specCH}) into 
\be\label{eq:specStr}
-f'' = z\, f\, \wt\omega
\ee
on the interval $(-1/2,1/2)$, where the measure $\wt\omega$ is given by
\[
  \int_{(0,\x]}d\wt\omega(s) = 4\int_{(0,x]}\cosh^2(s/2)d\omega(s).
\]
Notice that this change of variables also explains the coefficients in (\ref{eqnan}). In the case when $\wt\omega$ is a nonnegative measure, (\ref{eq:specStr}) describes small oscillations of an inhomogeneous string with mass density $\wt\omega$ and is known in the literature as {\em the string spectral problem}. Direct and inverse spectral theory for (\ref{eq:specStr}) with a nonnegative measure $\wt\omega$ was developed by Krein in the early 1950's  \cite{kakr74} and subsequently applied to study interpolation and filtration problems for stationary stochastic processes; see \cite{dymc76}. The first problem which appears in this context is {\em how to understand the differential equation~(\ref{eq:specStr}) with measure coefficients}? Krein suggested to replace the differential equation by an integral equation and as a result one can successfully develop the basic direct spectral theory for the string spectral problem (see, e.g., \cite{dymc76, kakr74}).
In fact, this can be done for a much larger class of Sturm--Liouville-type spectral problems \cite{MeasureSL}.

The inverse spectral problem aims to recover the coefficients of the differential equation (in our case the measure $\wt\omega$) from the spectral data. In order to explain the solution to the inverse spectral problem for strings, let us consider (\ref{eq:specStr}) on the interval $[0,L)$ assuming that $L\in(0,\infty]$ and $\wt\omega$ is a nonnegative measure on $[0,L)$. The pair $\cS_+=(L,\wt\omega)$ is called a string ($L$ and $\wt\omega$ are its length and mass distribution, respectively). For a given string $\cS_+$, let $c(z,\x)$ and $s(z,\x)$ be a fundamental system of solutions to (\ref{eq:specStr}) satisfying the following boundary conditions 
\[ 
c(z,0) = s'(z,-0) =1,\qquad c'(z,-0) = s(z,0)=0. 
\]
These solutions are entire functions in $z$ for every $x\in [0,L)$. Moreover, the limit 
\be\label{eq:Mstring}
M(z):=\lim_{\x\to L}-\frac{c(z,\x)}{z\, s(z,\x)}
\ee   
exists and is finite for all $z\in\C\setminus[0,\infty)$. 
The function $M:\C\setminus [0,\infty)\to \C$ is called {\em the Weyl--Titchmarsh function of the string $\cS_+$}. It turns out that the function $M$ is a Stieltjes function, that is, it is analytic on $\C\setminus [0,\infty)$, $M(z^\ast) =M(z)^\ast$ and $\im(z)\im\, M(z)\ge 0 $ for all $z\in\C\setminus [0,\infty)$ and $M(z)\ge 0$ for all $z<0$. Every Stieltjes function admits  a unique integral representation which, in our case, reads as follows
 \[
 M(z) = \wt\omega(\{0\}) - \frac{1}{L\, z} + \int_{(0,\infty)} \frac{d\rho(\lambda)}{\lambda - z}, \quad z\in\C\setminus\R,
 \] 
 where $\rho$ is a nonnegative measure on $(0,\infty)$ such that the integral 
 \[
 \int_{(0,\infty)}\frac{d\rho(\lambda)}{1+\lambda}
 \]
 is finite. The measure $\rho$ is called {\em the spectral measure of $\cS_+$} and it contains all the spectral information about $\cS_+$. Notice that (\ref{eq:Mstring}) defines a map from the set of all strings to the set of Stieltjes functions. A celebrated result of Krein states that this map is a bijection, that is, {\em every Stieltjes function $M$ is the Weyl--Titchmarsh function of a unique string $\cS_+$} (see \cite{dymc76, kakr74, kowa82}). Moreover, this map is homeomorphic with respect to appropriate weak topologies. The proof of this result is based on the Krein--de Branges theory of Hilbert spaces of entire functions and can be found in \cite{dymc76}. 

There were many attempts to extend the results of Krein to the case when $\wt\omega$ is a real-valued measure (see, e.g., \cite{bel87, be04,bebrwe08,bebrwe12, eck12, kau06}) but only insufficient partial results were available in this case, even though applications to the CH equation demand to have such a generalization. One of the key problems in this context is the fact that the natural framework for Krein strings is the Hilbert space $L^2([0,L);\wt\omega)$. However, the inner product is nonnegative precisely when the measure $\wt\omega$ is nonnegative. Otherwise $L^2([0,L);\wt\omega)$ becomes {\em a Krein space}, an indefinite inner product space \cite{azio89}. In this respect, the spectral problem~(\ref{eq:specStr}) is one of the basic toy models in the spectral theory of operators in Krein spaces (see \cite{fl96,fl14}). Notice that during the last three decades a lot of work has been devoted to the study of (\ref{eq:specStr}), due to its importance in numerous applications (we refer to \cite{fl96, fl14, ko13} for further information and references).    

Returning to the inverse spectral problem for indefinite strings, a first guess could suggest that instead of the class of Stieltjes functions one obtains the entire class of Herglotz--Nevanlinna functions.  
 However, this is not the case as it turned out that the class of spectral problems (\ref{eq:specStr}) with real-valued Borel measures $\wt\omega$ is too narrow in this respect, even for rational Herglotz--Nevanlinna functions (notice that in the multi-peakon situation there are cases when the inverse problem cannot be solved).  
 The deeper reason for why this fails in the real-valued case in some sense lies in the fact that the class of real-valued Borel measures is not closed with respect to a particular topology, whereas the class of nonnegative Borel measures is. 
 Altogether, it does not seem very likely that there is a simple and concise description of the class of Weyl--Titchmarsh functions that arise from the spectral problem  (\ref{eq:specStr}) with real-valued Borel measures $\wt\omega$. 

One way to overcome this problem by means of extending the class of spectral problems was suggested by Krein and Langer \cite{krla79}, who considered the modified differential equation  
 \be\label{eqnSPla}
  - f'' = z\, f\, \wt\omega + z^2\, f\,\wt\dip 
 \ee
 on an interval $[0,L)$, where $\wt\omega$ is a real-valued Borel measure on $[0,L)$ and $\wt\dip$ is a nonnegative Borel measure on $[0,L)$. 
 In particular, they showed in \cite{krla79} that indeed every rational Herglotz--Nevanlinna function arises as the Weyl--Titchmarsh function of such a spectral problem.
 However, the totality of all Weyl--Titchmarsh functions which are obtained in this way is still a proper subset of the class of Herglotz--Nevanlinna functions (see \cite{lawi98}). 
 
Let us consider the set of all triplets $\cS=(L,\wt\omega,\wt\dip)$, where $L\in(0,\infty]$, $\wt\omega\in H^{-1}_{\loc}[0,L)$ is a real-valued distribution on $[0,L)$ and $\wt\dip$ is a nonnegative Borel measure on $[0,L)$. The corresponding differential equation (\ref{eqnSPla}) has to be understood in a suitable distributional sense in general (see \cite{sash99} and also \cite{ecko16,CHPencil}). 
Following \cite{ecko16}, $\cS$ is called {\em an indefinite generalized string}. It turns out that every Herglotz--Nevanlinna function can be realized as the Weyl--Titchmarsh function of a unique generalized string. More precisely, under the above assumptions on the coefficients, the spectral problem~(\ref{eqnSPla}) admits a fundamental system of solutions $c(z,x)$ and $s(z,x)$ with similar properties as for Krein strings. Moreover, one can define the Weyl--Titchmarsh function via (\ref{eq:Mstring}) for all $z\in\C\setminus\R$, however, $M$ is now only a Herglotz--Nevanlinna function, which admits the  integral representation
\[
M(z) = \wt\dip(\{0\})\,z + c - \frac{1}{L\, z} + \int_{\R} \left( \frac{1}{\lambda-z} - \frac{\lambda}{1+\lambda^2}\right) d\rho(\lambda),\quad z\in\C\setminus\R,
\]
with some $c\in\R$ and a nonnegative Borel measure $\rho$ on $\R$ with $\rho(\lbrace0\rbrace)=0$ for which the integral  
\[
 \int_\R \frac{d\rho(\lambda)}{1+\lambda^2}
\]
is finite. The main result in \cite{ecko16} states that {\em every Herglotz--Nevanlinna function is the Weyl--Titchmarsh function of a unique generalized indefinite string}. The proof of this result is based on de Branges' solution of the inverse spectral problem for $2\times2$ canonical systems 
\cite{dB68}. Let us also mention that in view of applications to the CH equation, the regularity of coefficients is exactly what is desired.

 \section{Global conservative solutions}\label{sec:conserv} 

 Even though solutions of the CH equation (\ref{eqnCH}) may blow up in finite time, it turned out that it is always possible to continue them globally in a reasonable weak sense~\cite{xizh00}. Since such continuations are not unique anymore in general, one is led to impose additional constraints on them in order to guarantee uniqueness. Among all possible continuations,  there are two (in some sense) extremal cases; {\em dissipative} continuations and {\em conservative} continuations. Whereas the former one postulates a loss of energy due to wave-breaking, the latter one requires the total energy of the solution (measured by the $H^1(\R)$ norm) to be conserved. 
     
\begin{example}[Peakon-antipeakon interaction]\label{ex:4.1} Let us continue with Example \ref{ex:3.1}. Assume for simplicity that $p_1(0)=-p_2(0)>0$ and $q_1(0)=-q_2(0)<0$. Notice that then $p_1(t) = - p_2(t)$, $q_1(t) = - q_2(t)$ for all $t\in(0,t^\times)$, which, in particular, implies that $u(\ledot,t)$ is odd for all $t\in(0,t^\times)$. 
Taking into account the formulas (\ref{eq:system_2peaks})--(\ref{eq:Q_2peaks}), one can show that $u(x,t)\to 0$ for all $x\in\R$ as $t\to t^\times$. On the other hand, one finds that 
\be
\int_{q_1(t)}^{q_2(t)} |u(x,t)|^2 + |u_x(x,t)|^2\, dx = \frac{P(t)^2}{2}\left(1-\E^{-2Q(t)}\right) \to 4H_0^2=4H(p,q) 
\ee
as $t\to t^\times$ (to this end, use (\ref{eq:energy_2peaks}) and notice that $Q(t) =2q_2(t)\to 0$ as $t\to t^\times$). The latter shows that in the limit, the whole energy concentrates at a single point. 
\end{example} 

 In order for the Cauchy problems for these weak kinds of solutions to be well-posed, it is thus necessary to introduce an additional quantity $\mu$, which measures the energy density of the solution.  
  A solution now consists of a pair $(u,\mu)$, where $\mu$ is a nonnegative Borel measure, whose absolutely continuous part is determined by $u$ via 
 \be
  \mu_{ac}(B,t) = \int_B |u(x,t)|^2 + |u_x(x,t)|^2 dx, \quad t\in\R,
 \ee
 for each Borel set $B\subseteq\R$. 
 Within this picture, blow-up of solutions corresponds to concentration of energy (measured by $\mu$) to sets of Lebesgue measure zero. Existence of such global dissipative and conservative solutions for initial data in $H^1(\R)$ has been established in \cite{brco07, brco07a, hora07,hora07b, hora09} by means of a generalized method of characteristics that relies on a transformation from Eulerian to Lagrangian variables. Let us also stress the fact that, within the Lagrangian viewpoint, the CH flow is fine for all times was noticed in \cite{mck03}.

\section{Complete integrability}\label{sec:integrability} 

Although there is a multitude of further possibilities to guarantee uniqueness, the notion of {\em global conservative solutions} is the suitable one for our purposes, as it retains the completely integrable structure of the CH equation.  
 In fact, it turns out that this kind of solution indeed allows an associated isospectral problem. 
 Of course, an eligible modification of the isospectral problem~(\ref{eq:Spec}) now also has to incorporate the singular part of $\mu$ (with respect to the Lebesgue measure) in some way. 
 It will turn out that the appropriately generalized spectral problem is simply given by
\be \label{eqnGSP}
 -y'' + \frac{1}{4} y = z\, y\,\omega(\ledot,t)  + z^2\, y \,\upsilon(\ledot,t),
\ee
where $\upsilon$ denotes the singular part of the measure $\mu$. Using the Liouville transform (\ref{eq:LTstring}) it is not difficult to see that (\ref{eqnGSP}) is transformed to the following equation
\[
 -f'' = z\, f\, \wt{\omega} + z^2\, f\, \wt{\upsilon},
\]
on the interval $(-1/2,1/2)$. 
The idea for considering this particular spectral problem goes back to work of Krein and Langer \cite{krla79} on the indefinite moment problem and generalized strings, which carry not only negative mass but also dipoles. It is indeed interesting to observe the parallels between the developments described here in connection with conservative solutions of the CH equation and their work from the 1970s.

\begin{example}\label{ex:5.1}
Let us continue with Example \ref{ex:4.1}. We observed in Section \ref{sec:MP} that the multi-peakon flow can be solved by employing the solution of the Stieltjes moment problem. Let $M(\ledot,t)$ denote the Weyl--Titchmarsh function of the corresponding spectral problem at time $t\in[0,t^\times)$. Using (\ref{eqnContFrac}) and (\ref{eqnan}), one then observes that
\[
zM(z,t) \to zM(z,t^\times):= 1+\frac{1}{\displaystyle -1/2 + \frac{1}{\displaystyle16 H_0^2\, z^2 + \frac{1}{-1/2 }}}
=\frac{4H_0^2z^2}{1-4H_0^2z^2},
\]
for all $z\in \C\setminus\R$ as $t\to t^\times$. 
 First of all, note that the function $M(\ledot,{t^\times})$ is a Herglotz--Nevanlinna function. Moreover, the function $M(\ledot,{t^\times})$ is the Weyl--Titchmarsh function (in the sense of Section \ref{sec:MP}) for the quadratic spectral problem
\[
-y'' + \frac{1}{4} y =z^2\,\upsilon\, y,
\]
where $\upsilon = 4H_0^2\, \delta_0$.
\end{example}

\subsubsection*{Conservative multi-peakons}

A detailed description of global conservative multi-peakon solutions was given in \cite{hora07b}. Following \cite{hora07b}, a global conservative solution $(u,\mu)$ of the CH equation is said to be {\em a multi-peakon solution} if for some $t_0\in\R$, the measure $\mu(\ledot,t_0)$ is absolutely continuous and the function $u(\ledot,t_0)$ is of the form (\ref{eqnMP}). For these solutions, at any time $t\in\R$, the quantities $\omega$ and $\dip$ are of the form 
\be\label{eq:coeffMP}
\omega(\ledot,t) = \sum_{n=1}^{N(t)} \omega_n(t)\delta_{x_n(t)}, \quad\quad \dip(\ledot,t) = \sum_{n=1}^{N(t)} \dip_n(t) \delta_{x_n(t)},
 \ee
where $N(t)\in\N_0$,  $x_1(t),\ldots,x_{N(t)}(t)\in\R$ are strictly increasing,  $\omega_n(t)\in\R$ and $\dip_n(t)\geq 0$  for $n=1,\ldots,N(t)$. In this case, the spectral problem (\ref{eqnGSP}) can be treated similarly to the multi-peakon case (see Section \ref{sec:MP}) with the only modification concerning the norming constants (cf.\ (\ref{eq:NC})) associated with an eigenvalue $\lambda$:
\be\label{eq:NCmod}
  \frac{1}{\gamma_{\lambda}(t)}  := \int_\R |\phi_-'(\lambda,x,t)|^2 dx+\frac{1}{4}\int_\R |\phi_-(\lambda,x,t)|^2 dx + \int_\R |\lambda\phi_-(\lambda,x,t)|^2 d\dip(x,t).
  \ee 
The first two trace formulas (see \cite{ecko14})
\[
\sum_{\lambda\in\sigma} \frac{1}{\lambda}  = \int_\R d\omega, \qquad   \sum_{\lambda\in\sigma} \frac{1}{\lambda^2}  =  2 \int_\R d\mu,
\]
indicate that (\ref{eqnGSP}) might serve as an isospectral problem for the conservative CH flow (compare with (\ref{eq:H})). And indeed, the main result in \cite{ecko14} states that {\em the pair $(u,\mu)$ is a global conservative multi-peakon solution of the CH equation if and only if the problems (\ref{eqnGSP}), (\ref{eq:coeffMP}) are isospectral with}  
  \be\label{eqnTE}
   \dot{\gamma}_{\lambda} =  \frac{1}{2\lambda} \gamma_{\lambda}, \quad  \lambda\in\sigma. 
  \ee
Moreover, utilizing the solution of the indefinite moment problem given by  Krein and  Langer \cite{krla79}, it was proved in \cite{ecko14} that the conservative CH flow is completely integrable by the inverse spectral transform in the multi-peakon case. Let us also mention that the Krein--Langer solution of the indefinite moment problem plays the same role for the conservative multi-peakon flow as the solution to the Hamburger moment problem for the finite Toda lattice on the line (see \cite{mos75}).

\subsubsection*{General conservative solutions}

In order to be able to treat general conservative solutions, we need to allow $u$ to be an arbitrary function in $H^1(\R)$, rendering $\omega$ a distribution in $H^{-1}(\R)$, and $\upsilon$ to be a nonnegative finite  Borel measure on $\R$. 
Thus, the differential equation (\ref{eqnGSP}) has to be understood in a suitable distributional sense in general (see \cite{sash99} and also \cite{ecko16,CHPencil}). 
The associated spectrum is then again defined as the set of all those $z\in\C$ for which there is a solution of the differential equation~(\ref{eqnGSP}) that belongs to $H^1(\R)$. 

For a given global conservative solution $(u,\mu)$ of the CH equation, one can indeed show that the spectrum associated with~(\ref{eqnGSP}) is  independent of time. 
Under the condition that the associated spectrum $\sigma$ is a discrete set of nonzero reals that satisfies 
\be \label{eqnTC}
 \sum_{\lambda\in\sigma} \frac{1}{|\lambda|}  < \infty,
\ee
one may prove existence of two (unique) solutions $\phi_\pm(z,\redot,t)$ of the differential equation~(\ref{eqnGSP})  with the spatial asymptotics
\[
 \phi_\pm(z,x,t) \sim \E^{\mp\frac{x}{2}}, \qquad x\rightarrow\pm\infty,
 \]
for every $z\in\C$ and $t\in\R$, generalizing the corresponding solutions from the multi-peakon case.  These solutions are entire of genus zero when considered as a function of $z\in\C$. The asymptotic normalization implies that
the solutions $\phi_\pm(z,\redot,t)$ are square integrable near $\pm\infty$, but will not be square integrable near $\mp\infty$ in general. 
However, we see that some $z\in\C$ belongs to the spectrum $\sigma$ if and only if the functions $\phi_-(z,\redot,t)$ and $\phi_+(z,\redot,t)$ are linearly dependent. Thus for every $\lambda\in\sigma$ we may write 
\be
\phi_+(\lambda,x,t) = c_\lambda(t) \phi_-(\lambda,x,t), \quad x,\,t\in\R,
\ee
for some function $c_\lambda$. 
 As in the case of multi-peakons in the previous section, it turns out that the time evolution of these spectral quantities and the norming constants is linear and simply given by  
\[
 \dot{\gamma}_\lambda =  \frac{1}{2\lambda} \gamma_\lambda, \qquad \dot{c}_\lambda = -\frac{1}{2\lambda} c_{\lambda}, 
\]
for every $\lambda\in\sigma$. 
 Since one can show that the pair $(u,\mu)$ is uniquely determined by the spectrum $\sigma$ and the corresponding sequence of norming constants $\{\gamma_\lambda\}_{\lambda\in\sigma}$, this implies that the transformation $(u,\mu)\mapsto\{\gamma_\lambda\}_{\lambda\in\sigma}$ maps the conservative CH flow on isospectral sets to a simple, explicitly solvable linear flow on $\R_+^\sigma$. 
This is reminiscent of the fact that the conservative CH flow can be viewed as a completely integrable infinite dimensional Hamiltonian system.

\section{Long-time asymptotics}\label{sec:asymp}

In the theory of linear partial differential equations a key role is played by dispersion which
implies that waves decay over time as different plane waves travel at different speeds leading
to destructive interference. On the other hand, nonlinear partial differential equations
lead to wave breaking and it thus came as a surprise when John Scott Russell in 1834 observed
his famous {\em Wave of Translation} in a narrow channel. At that time his observation
seemed to contradict generally accepted believes and it took another 37 years until
Boussinesq (1871), Lord Rayleigh (1876) and finally Korteweg and his student
de Vries (1895) supported his observation with mathematical theory. Despite these results,
the real significance of solitons remained hidden for another century until Zabusky and Kruskal
in 1965 used the KdV equation (\ref{eqnKdV}) to explain the Fermi--Pasta--Ulam experiment
(another observation --- this time in one of the first computer experiments --- which originally seemed to contradict
generally accepted believes). In fact, their numerical experiments suggested that any decaying initial wave profile
of the KdV equation asymptotically splits into a finite number of solitons plus a dispersive tail. Hence the solitons
were rendered from a peculiar solution into a central object. Moreover, many other integrable wave equations with
soliton solutions were found and proving that decaying initial conditions asymptotically split into a sum of
solitons over time became a major task known as soliton resolution conjecture (see \cite{tao} for a review).

Building on the inverse scattering transform discovered by Gardner, Greene, Kruskal and Miura,
it was possible to give a detailed description of the long-time asymptotics for the KdV equation with decaying initial
conditions. While first approaches required an ansatz for the asymptotic form of the solution, this was eventually
overcome by Its (1981) who outlined how to complete an original idea by Manakov (1974). Finally, Deift and
Zhou (1993) turned these ideas into a fully rigorous theory now known as nonlinear steepest descent method for
oscillatory Riemann--Hilbert problems. We refer to \cite{gt} for further historic details and an expository introduction to this method.
See also the review \cite{its} for further information.

This method also applies to the CH equation, but only when the constant $\kappa$ and the momentum $\omega$ are strictly positive.
Since this method is rather technical, even a brief overview is beyond the scope of this review and we refer to \cite{bkst} for further details.
Here we want to focus on the case $\kappa=0$ when there is no dispersive tail and the solution splits into a pure (but infinite)
sum of peakons. Despite the fact that this behavior was already expected from the original work of Camassa and Holm and
emphasized as an important conjecture by McKean in~\cite{mck03}, this corresponding peakon resolution conjecture for the
CH equation remained open and was only solved by two of us in~\cite{IsospecCH}. As already mentioned, the standard techniques
developed so far did not apply in this situation and a new approach was required. Here we will sketch a particularly
simple method based on a novel coupling problem for entire functions \cite{CouplProb}.

For this purpose, suppose that $(u,\mu)$ is a global conservative solution of the CH equation. 
As always the starting point is the isospectral problem (\ref{eqnGSP}), where we assume that the underlying spectrum satisfies the condition~(\ref{eqnTC}). 
 With the principal notation from the previous section, we now proceed with a simple rescaling
\be
  \Phi_\pm(z,x,t) = \E^{\pm\frac{x}{2}} \phi_\pm(z,x,t), \quad z\in\C, 
\ee
such that the coupling condition reads
\be
  \Phi_+(\lambda,x,t) = \E^{x-\frac{t}{2\lambda}} c_\lambda(0) \Phi_-(\lambda,x,t), \quad \lambda\in\sigma,
\ee
and switch to a moving frame, letting $t\to\infty$ as $\eta:=\frac{t}{2x}$ is kept constant.
Now for every $\lambda\in\sigma$ with $\eta^{-1}>\lambda^{-1}$ we have that $\E^{x-\frac{t}{2\lambda}} = \E^{\frac{t}{2}(\frac{1}{\eta}-\frac{1}{\lambda})}\to 0$ as $t\to\infty$
and for every $\lambda\in\sigma$ with $\eta^{-1}<\lambda^{-1}$ we have that $\E^{x-\frac{t}{2\lambda}} = \E^{\frac{t}{2}(\frac{1}{\eta}-\frac{1}{\lambda})}\to \infty$ as $t\to\infty$.
Consequently, in the first case the coupling condition asymptotically reads $\Phi_+(\lambda,x,\infty)=0$ and in the second case $\Phi_-(\lambda,x,\infty)=0$.
Finally, one uses the fact that the function
\be
  \frac{z\, \Phi_-(z,x,t) \Phi_+(z,x,t)}{W(z)}, \quad z\in\C\backslash\R,
\ee
is the diagonal of the kernel of the Green's function of our isospectral problem and as such, a meromorphic Herglotz--Nevanlinna function. But it is a well-known result
that for a meromorphic Herglotz--Nevanlinna function the poles and zeros are interlacing implying that in the limit as $t\to\infty$ the zeros of $\Phi_-(z,x,t) \Phi_+(z,x,t)$
will cancel with the zeros of $W(z)$, except for at most one in the case $\eta\in\sigma$. Hence we can cancel these zeros from the picture and are asymptotically left
with a coupling problem which has at most one coupling condition. 
Since such a problem can easily be solved explicitly, this leads to the following
asymptotics (see \cite{CouplProb})
 \be
  u(x,t) = \sum_{\lambda\in\sigma} \frac{1}{2\lambda} \E^{-\left|x-\frac{t}{2\lambda} + \xi_\lambda\right|} + \oo(1),
 \ee
which hold uniformly for all $x\in\R$ as $t\rightarrow\infty$. The phase shifts $\xi_\lambda$ appearing in this formula are given by
 \be
   \xi_\lambda = \ln|c_{\lambda}(0)| + \sum_{k\in\sigma\backslash\lbrace\lambda\rbrace} \sgn\left( \frac{1}{\lambda} - \frac{1}{k} \right) \ln\left|1-\frac{\lambda}{k}\right|, \quad \lambda\in\sigma.
 \ee
 Therefore, the typical long-time behavior of the function $u$ of our global conservative solution~$(u,\mu)$ of the CH equation can be depicted as follows:
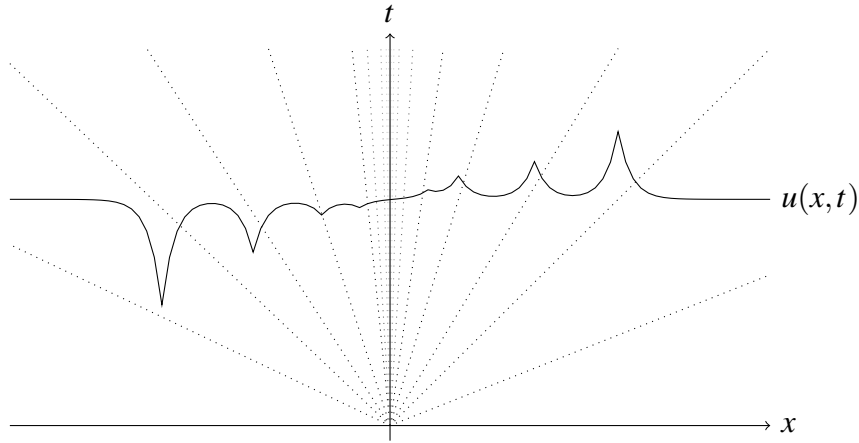
\begin{center}
 \begin{tikzpicture}[domain=-1:9, samples=101]
\draw[color=black] plot (\x,{ 3 + 0.9*exp(-6*abs(\x-7.)) + 0.5*exp(-6*abs(\x-5.9)) + 0.3*exp(-6*abs(\x-4.9)) - 1.4*exp(-6*abs(\x-1)) - 0.7*exp(-6*abs(\x-2.2)) - 0.2*exp(-6*abs(\x-3.1))+ 0.1*exp(-6*abs(\x-4.5)) - 0.1*exp(-6*abs(\x-3.6)) }) node[right] {$u(x,t)$};

\draw[dotted] (4,0) -- (9,2) node[right] {};
\draw[dotted] (4,0) -- (9,5) node[right] {};
\draw[dotted] (4,0) -- (7,5) node[right] {};
\draw[dotted] (4,0) -- (5.5,5) node[right] {};
\draw[dotted] (4,0) -- (4.7,5) node[right] {};
\draw[dotted, color=black!60] (4,0) -- (4.3,5) node[right] {};
\draw[dotted, color=black!40] (4,0) -- (4.12,5) node[right] {};
\draw[dotted, color=black!20] (4,0) -- (4.05,5) node[right] {};
\draw[dotted, color=black!10] (4,0) -- (4.02,5) node[right] {};

\draw[dotted] (4,0) -- (-1,2.4) node[left] {};
\draw[dotted] (4,0) -- (-1,4.8) node[left] {};
\draw[dotted] (4,0) -- (0.8,5) node[left] {};
\draw[dotted] (4,0) -- (2.4,5) node[left] {};
\draw[dotted] (4,0) -- (3.5,5) node[left] {};
\draw[dotted, color=black!60] (4,0) -- (3.7,5) node[left] {};
\draw[dotted, color=black!40] (4,0) -- (3.88,5) node[left] {};
\draw[dotted, color=black!20] (4,0) -- (3.95,5) node[left] {};
\draw[dotted, color=black!10] (4,0) -- (3.98,5) node[left] {};

\draw[->] (-1,0) -- (9,0) node[right] {$x$};
\draw[->] (4,-0.2) -- (4,5.2) node[above] {$t$};
\end{tikzpicture}
\end{center}
Each of the dotted lines $x= \frac{1}{2\lambda}t$ emanating from the origin and accumulating towards the $t$-axis corresponds to an eigenvalue $\lambda\in\sigma$ of the underlying isospectral problem.  After long enough time, one can see that the solution $u$ splits into a train of single peakons, each of which travels along one of the rays, with height and speed determined by the corresponding eigenvalue.

\bigskip

\bigskip\bigskip
\noindent
{\bf\large Acknowledgements}.
Research supported by the Austrian Science Fund (FWF) under Grants No.\ J3455 and P26060.


\vfill

{\it Adresse der Autoren:
Fakult\"at f\"ur Mathematik, Universit\"at Wien,
Oskar-Morgenstern-Platz 1, A 1090 Wien.
}

\Artikelende

\end{document}